# Theoritical Kinetic Study of the Ring Opening of Cyclic Alkanes

*B. Sirjean[1], P.A. Glaude[1], M.F. Ruiz-Lopez[2], R. Fournet[1*]*

[1] Département de Chimie-Physique des Réactions, UMR CNRS-INPL, Nancy Université, Nancy, France
[2] Equipe de Chimie et Biochimie Théorique, UMR CNRS-UHP, Nancy Université, Vandoeuvre-les-Nancy, France

## Abstract
This work reports a theoretical study of the gas phase unimolecular decomposition of cyclobutane, cyclopentane and cyclohexane by means of quantum chemical calculations. A biradical mechanism has been envisaged for each cycloalkane, and the main routes for the decomposition of the biradicals formed have been investigated at the CBS-QB3 level of theory. Thermochemical data ($\Delta H°_f$, S°, $C°_p$) for all the involved species have been obtained by means of isodesmic reactions. Rate constants have been derived for all elementary reactions using transition state theory at 1 atm and temperatures ranging from 600 to 2000 K. These values have been compared with the few data available in the literature and showed a rather good agreement.

## Introduction
Among fuels components, cyclic hydrocarbons, particularly cycloalkanes, represent an important class of compounds. These molecules are produced during the gas-phase processes though they can also be present in the reactants in large amounts; for example, a commercial jet fuel contains about 26% of naphtenes and condensed naphtenes, while in a commercial diesel fuel, this percentage reaches 40 % [1]. During combustion or pyrolysis processes, cycloalkanes can lead to the formation of toxics or soot precursors. Several experimental and modeling studies have been carried out on the oxidation of cyloalkanes in gas phase [2-8]. However, the modeling of their combustion remains difficult due to the low number of kinetic and thermodynamic data for the relevant species.

Thermal decomposition of cyclobutane has been experimentally studied and rate constants for the ring opening yielding two ethylene molecules have been measured [9-12]. Theoretical studies on cyclobutane have mainly focused on the reverse reaction of cycloaddition of two ethylene molecules [13-16]. Moreover, Pedersen et al. [17] have showed the validity of the biradical hypothesis by direct femtosecond studies of the transition-state structures. Theses studies have highlighted the fact that cycloaddition of two ethylene molecules can proceed through a tetramethylene biradical intermediate or a concerted reaction that directly leads to cyclobutane formation. However, the latter reaction has been shown to have a high activation energy due to steric effects and to be much less favorable than the biradical process.

The study of ring opening of cyclobutane is interesting from a theoretically point of view, but larger cycloalkanes such as cyclopentane or cyclohexane are mainly involved in fuels. In spite of that, the unimolecular initiation of these molecules has been little investigated. Ring opening of cyclopentane and cylohexane has been experimentally studied by Tsang [18,19] who reported the main routes of decomposition of these molecules. For cyclohexane, a pathway leading to the formation of 1-hexene has been considered only whereas for cyclopentane, the processes leading to either 1-pentene or to cyclopropane + ethylene have both been investigated, in accordance with the products experimentally detected. In addition, Tsang [18] has shown that the experimental global rate parameters are consistent with a biradical mechanism for ring opening:

The present work analyzes the ring opening of cyclobutane, cyclopentane, and cyclohexane by means of high level quantum chemical calculations in order to obtain accurate rate constants for elementary reactions. We explore several plausible pathways that could be involved in the decomposition of the biradicals.

## Computational method
Quantum chemical computations were performed on an IBM SP4 computer with the Gaussian03 software package [20]. The high-level composite method CBS-QB3 has been used. Analysis of vibrational frequencies confirmed that all transition structures (TS) have one imaginary frequency. Intrinsic Reaction Coordinate (IRC) calculations have been systematically performed at the B3LYP/6-31G(d) level to ensure that the TSs connect the desired reactants and products.

Only singlet states have been considered for the biradicals and their study at the composite CBS-QB3 level requires two specific comments. First, geometry optimization is performed at the density functional theory (DFT) level using an unrestricted one-determinantal wavefunction (UB3LYP/CBSB7 method). Though such an approach might be questionable for describing open-shell singlet biradicals, previous studies have shown that the obtained geometries compare well with those predicted at more refined computational levels [21]. Second, owing to strong singlet-triplet mixing, spin contamination in singlet biradical calculations is quite large and the computed energy requires some correction. The sum method [22] has usually been employed but cannot be directly applied within the composite CBS-QB3 approach. In the latter case, a correction for spin-contamination is already added to the total energy: This correction has been derived for systems displaying a small spin-contamination and is expected to work properly for a triplet. However, for a singlet biradical, the standard formula is not suitable and leads to a systematic error in the CBS-QB3 energy correction (of about 6 kcal.mol$^{-1}$). A specific approach with a new parameter has been determined to allow the calculation of singlet biradicals that is described extensively elsewhere [23,24].

[*] Corresponding author : Rene.Fournet@ensic.inpl-nancy.fr
Associated Web site: http://www.ensic.inpl-nancy.fr/DCPR/
DCPR-ENSIC, BP 20451
1, rue Grandville
54001 NANCY Cedex

## Thermochemical data

Thermochemical data for all the species involved in this study have been derived from energy and frequency calculations. In the CBS-QB3 method, harmonic frequencies, at the B3LYP/cbsb7 level of theory, are scaled by a factor 0.99. Explicit treatment of the internals rotors has been performed with the *hinderedRotor* option of Gaussian03. A systematic analysis of the results obtained was made in order to ensure that internals rotors were correctly treated. Thus, for biradicals a practical correction was introduced in order to take into account the symmetry number of 2 for each $CH_2(\bullet)$ terminal group. This symmetry is not automatically recognized by Gaussian in the case of a radical group. In transition states, the constrained torsions of the cyclic structure have been treated as harmonic oscillators and the free alkyl groups as hindered rotations.

Enthalpies of formation of species have been calculated using isodesmic reactions excepted for cyclanes and 1-alkene for which more accurate experimental values [25] can be found in the literature. Several isodesmic reactions have been considered for the calculation of $\Delta H_f°$ to obtain an average value.

The computed entropy of cyclopentane has been corrected in order to take into account the experimental symmetry of the molecule ($D_{5h}$). Indeed, calculations lead to a non planar geometry with $C_1$ symmetry. In addition, a low frequency of 22 cm$^{-1}$ can be associated with a puckering motion of the ring. According to Benson [26], this pseudo-rotation is so fast that cyclopentane can be treated as dynamically flat (with a symmetry number $\sigma$=10). The computed value, 70.0 cal.mol$^{-1}$.K$^{-1}$, is in good agreement with the experimental value [26].

## Kinetic calculations

The rate constants involved in the mechanisms were calculated using TST [27]:

$$k_{uni} = \text{rpd}\,\kappa(T)\frac{k_b T}{h}\exp\left(\frac{\Delta S^{\neq}}{R}\right)\exp\left(-\frac{\Delta H^{\neq}}{RT}\right)$$

where $\Delta S^{\neq}$ and $\Delta H^{\neq}$ are the entropy and enthalpy of activation and rpd is the reaction path degeneracy. For reactions involving H-transfer, a transmission coefficient has been calculated in order to take into account tunneling effect [27]:

The calculation of $\Delta H^{\neq}$ was calculated from electronic energies of reactants, products, and TSs but also isodesmic enthalpy of reaction by taking the average value of the activation enthalpies calculated for the direct and reverse reactions in order to minimize systematic errors and to ensure that kinetics fit the thermochemical data.

The kinetic data are obtained with a fitting in the temperature range 600-2000K, with the following modified Arrhenius form:

$$k = A\,T^n \exp\left(-\frac{E_a}{RT}\right)$$

## Cyclobutane

Scheme 1 presents the mechanism obtained for the ring opening of cyclobutane. In this scheme, we only considered conformers of lowest Gibbs free energies, i.e. the gauche conformer in the case of $\bullet C_4H_8\bullet$.

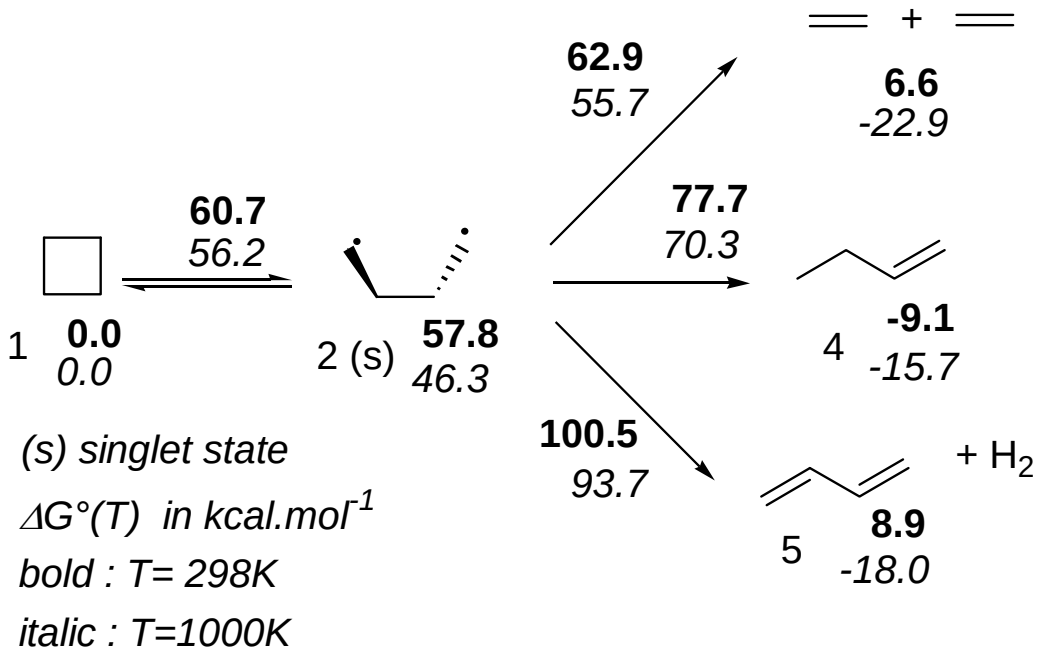

Scheme 1: Mechanism obtained for the ring opening of cyclobutane and for conformers of lowest energies.

The ring opening of cyclobutane involves a low Gibbs free energy of activation; $\Delta G^{\neq}$ = 60.7 kcal.mol$^{-1}$ at 298 K which corresponds to an activation enthalpy ($\Delta H^{\neq}$) of 62.7 kcal.mol$^{-1}$. This value is lower than that involved in the dissociation of a C-C bond in the corresponding linear alkane (≈86.3 kcal.mol$^{-1}$ [28]) because of the recovery of the ring strain energy in the cyclobutane molecule. Following ring opening, three ways of decomposition for tetramethylene have been investigated. The route leading to the formation of two ethylene molecules corresponds to the most favorable one. The process leading to 1-butene represents another possible pathway though this elementary reaction involves a stressed cyclic transition state for H-transfer and displays a high activation energy (17.6 kcal.mol$^{-1}$ vs 2.8 kcal.mol$^{-1}$ for the β-scission leading to $C_2H_4$). However, $C_4H_8$ has been identified as a minor product in experimental studies of cyclobutane decomposition [11]. The third route consists of the abstraction of two hydrogen atoms to form buta-1,3-diene and $H_2$. The reaction step has a high activation barrier (40 kcal.mol$^{-1}$) and should not compete with the previous mechanisms.

Table 1 gives the kinetics parameters for all the processes involved in Scheme 1.

Table 1: Rate parameters for the unimolecular initiation of cyclobutane at 1 atm, 600 ≤ T (K) ≤ 2000 K and related to scheme 1, in kcal, mol, s, K, units.

| | $k_{1-2}$ | $k_{2-1}$ | $k_{2-C2H4}$ | $k_{2-4}$ | $k_{2-5}$ |
|---|---|---|---|---|---|
| log A | 18.53 | 12.21 | 7.32 | 5.57 | 2.23 |
| n | -0.797 | -0.305 | 1.443 | 2.171 | 2.995 |
| $E_a$ | 64.85 | 1.98 | 3.03 | 16.44 | 37.61 |

As mentioned previously, gauche/trans conversion of (2) has been neglected in Scheme 1. However, the β-scission reaction involves a low $\Delta G^{\neq}$ (5.1 kcal.mol$^{-1}$ at 298K) that could be of the same order of magnitude than the rotational barrier around the central C-C bond. Accordingly, it can be interesting to consider the two conformations of the biradical, as detailed in Scheme 2.

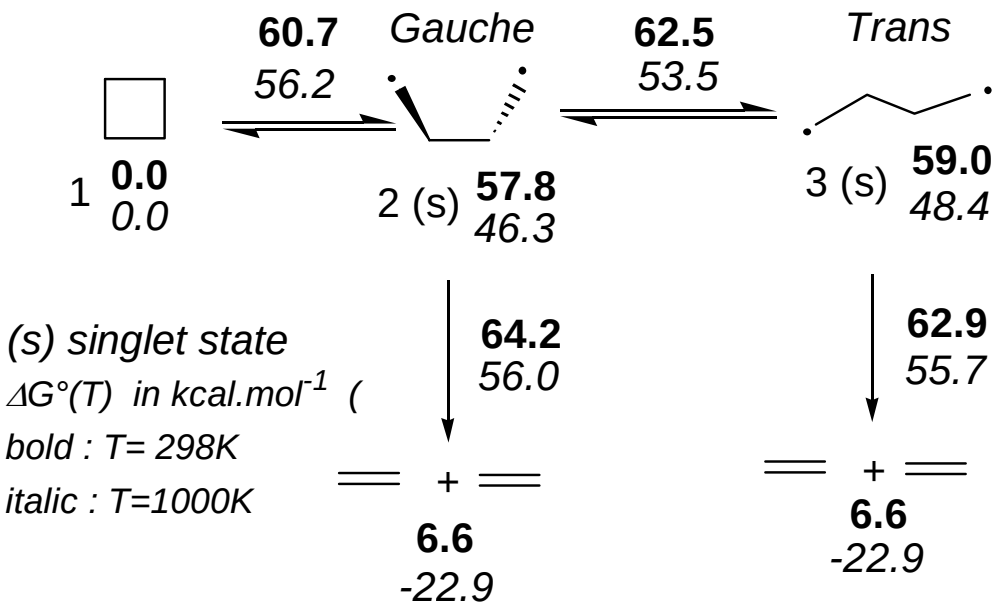

Scheme 2: Detailed mechanism of $C_2H_4$ formation from the ring opening of cyclobutane and obtained by considering different conformers of $C_4$ biradicals.

Cyclobutane ring-opening leads to the gauche biradical (2), which can decompose in two $C_2H_4$ molecules or rotate to give the trans conformer (3), which can also decompose in two $C_2H_4$ molecules. The kinetic parameters for all processes have been calculated. In order to validate our results, we compared the global rate constant for the process:

$$\text{Cyclobutane} \rightarrow C_2H_4 + C_2H_4,$$

measured experimentally [9, 10] with that derived from our computations (Schemes 1 and 2). We considered the quasi-stationary state approximation (QSSA) on the biradicals that allowed to calculate the global rate constants presented in table 2.

Table 2: Rate parameters for the global reaction $cC_4H_8 \rightarrow 2\ C_2H_4$, at P=1 atm, 600 ≤ T (K) ≤ 2000 K and related to Schemes 1 and 2, in kcal, mol, s, K, units.

| | $k_g^{scheme1}$ | $k_g^{scheme2}$ |
|---|---|---|
| log A | 20.29 | 21.52 |
| n | -1.259 | -1.606 |
| $E_a$ | 67.69 | 68.16 |

Figure 1 compares our results with experimental measurements for the decomposition of cyclobutane in two ethylene molecules. Computed rate constants are slightly lower than those obtained by Lewis et al. [10] (maximum factor 2.4) or by Barnard et al. [9] (maximum factor 4.3). It is interesting to note that differences between rates from both schemes decrease with temperature, which is consistent with rotational hindrance (20% at 600K and 8% at 2000 K). Though these differences are weak, the rate constant obtained by explicit consideration of the two conformations of the biradical (trans and gauche) is closer to experimental results than the rate constant calculated from Scheme 1.

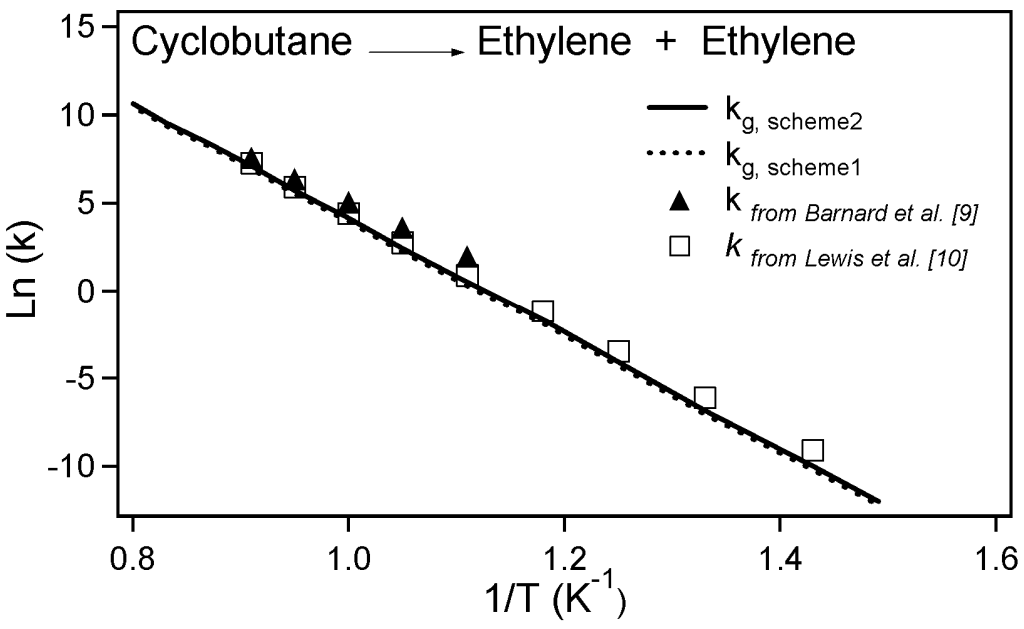

Figure 1: Calculated and experimental rate constants for the global reaction $cC_4H_8 \rightarrow C_2H_4 + C_2H_4$

**Cyclopentane**

Scheme 3 shows the global mechanism obtained for the ring opening of cyclopentane. As for cyclobutane, the global mechanism does not take into account the different conformations of the biradical $\bullet C_5H_{10}\bullet$ but only the conformation with the lowest Gibbs free energy.

Due to weaker ring strain energy, the Gibbs free energy of activation of the ring opening of cyclopentane ($\Delta G^{\neq}$ = 80.5 kcal.mol$^{-1}$) is higher than that obtained in cyclobutane ($\Delta G^{\neq}$ = 60.7 kcal.mol$^{-1}$).

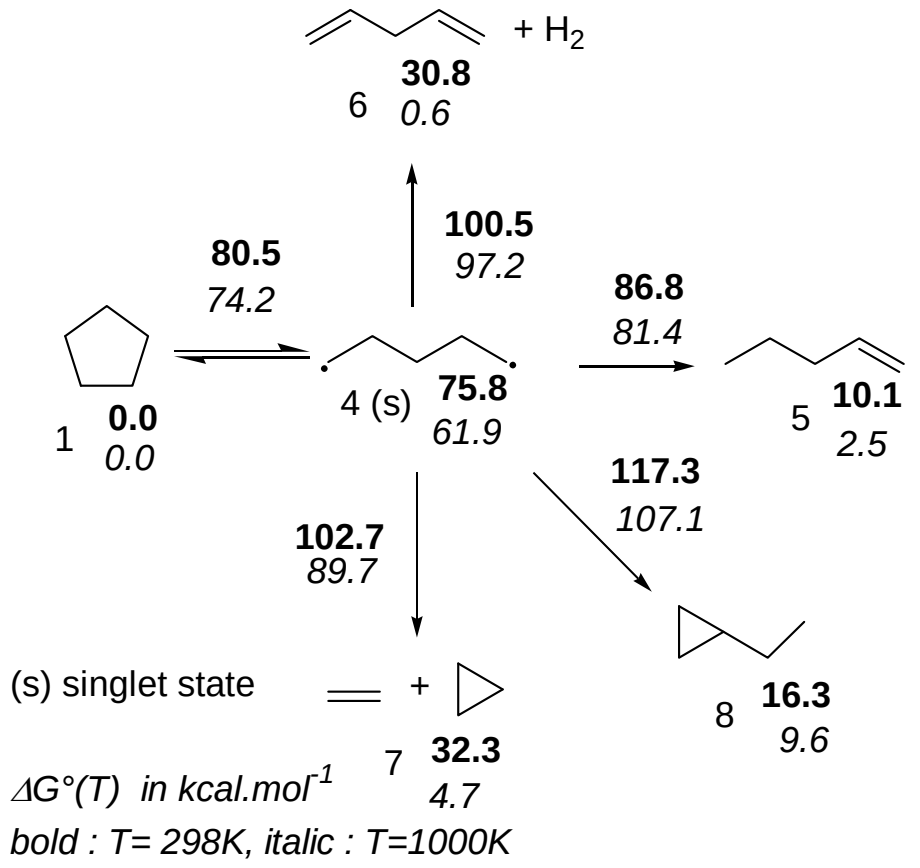

Scheme 3. Mechanism obtained for the ring opening of cyclopentane and for conformers of lowest energies.

Four routes of decomposition have been investigated for the biradical 4, the most favorable one being the formation of 1-pentene by H-transfer. This last reaction exhibits an activation energy much lower than the equivalent process in cyclobutane (6 kcal.mol$^{-1}$ *vs* 17.6 kcal.mol$^{-1}$ for $cC_4H_8$). This difference can be ascribed to different strain energy in the corresponding TSs.

Conversely, the β-scission process leading to ethylene and cyclopropane is much more difficult than that leading to ethylene in cyclobutane. In this last case, the presence of two radical centers in (•$C_4H_8$•) in β position weakens the inner C-C bond. Another point concerns the dehydrogenation reaction of biradical 4 leading to penta-1,4-diene (6) and $H_2$. At low temperature (298K), this reaction is competitive with β-scission (reaction 4→7) but becomes unimportant at high temperature (1000K) due to a low change of entropy between the biradical and TS. Finally, the reaction of biradical (4) to yield ethyl-cyclopropane (8) is unlikely to occur because it involves a high stressed cyclic transition state for H-abstraction and always displays a high activation energy. Table 3 gives the kinetics parameters for all the elementary processes involved in Scheme 3.

Table 3. Rate parameters for the unimolecular initiation of cyclobutane at P=1 atm, 600 ≤ T (K) ≤ 2000 K and related to Scheme 3, in kcal, mol, s, K, units.

| | $k_{1\text{-}4}$ | $k_{4\text{-}1}$ | $k_{4\text{-}5}$ | $k_{4\text{-}6}$ | $k_{4\text{-}7}$ | $k_{4\text{-}8}$ |
|---|---|---|---|---|---|---|
| log A | 18.11 | 9.89 | 6.77 | 0.51 | 9.78 | -3.07 |
| n | -0.466 | 0.311 | 1.480 | 3.015 | 1.1 | 4.157 |
| $E_a$ | 85.18 | 1.7 | 7.76 | 17.78 | 26.16 | 32.43 |

In Scheme 3, biradicals 4 represents the most stable conformer although it does not lead directly to the formation of 1-pentene. Since some activation energies, such as that of the reaction 4→5 (6.7 kcal.mol$^{-1}$), have values close to that of the rotation barrier heights, the role of rotational hindrance has been examined in the case of cyclopentane. Scheme 4 contains the detailed mechanism. Two conformations of the biradical •$C_5H_{10}$• are involved but only one (biradical 3) leads to the formation of 1-pentene. This result has been validated by IRC calculations.

It is worth noting the different role played by rotational hindrance in cyclopentane and cyclobutane. Indeed, in cyclobutane (Scheme 2), the rotational barrier is of the same order of magnitude as the activation energy for β-scission and both conformers may lead to the formation of ethylene molecules. In cyclopentane, the rotational barrier is lower than the activation energy for 1-pentene formation but only one conformation (biradical 3) allows the latter process.

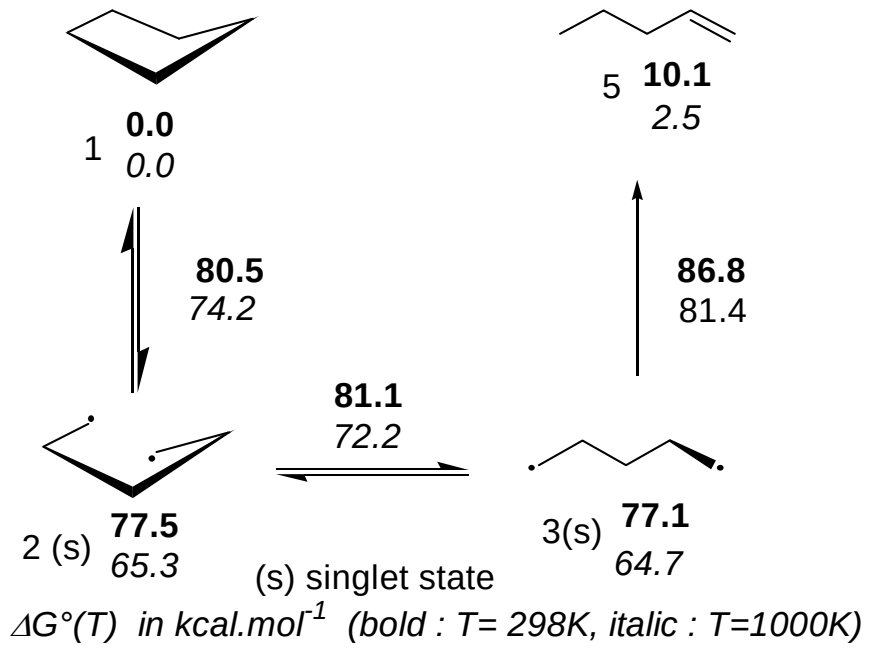

Scheme 4. Detailed mechanism of 1-pentene formation from the ring opening of cyclopentane and obtained by considering different conformers of $C_5$ biradicals.

In order to validate the rate parameters and to quantify the effect of the rotational barriers between biradicals 2 and 3, we have compared the global rate constants obtained experimentally by Tsang [18] with global rate constants estimated by QSSA performed on biradical 4 (Scheme 3) and biradicals 2 and 3 (Scheme 4). The kinetic parameters are given in Table 4.

Table 4: Rate parameters for the global reactions $cC_5H_{10}$→1-pentene and $cC_5H_{10}$→$cC_3H_6$ +$C_2H_4$ at 1 atm, 600 ≤ T (K) ≤ 2000 K and related to Schemes 3 and 4.

| | $k_{g(1-5)}^{scheme3}$ | $k_{g(1-7)}^{scheme3}$ | $k_{g(1-5)}^{scheme4}$ |
|---|---|---|---|
| log A (s$^{-1}$) | 20.39 | 24.33 | 20.06 |
| n | -0.970 | -1.542 | -0.878 |
| $E_a$ (kcal/mol) | 92.86 | 112.49 | 92.23 |

Comparison of calculated and experimental rate constants are presented in Figure 2. In the range 1000K –1200K, the calculated rate constants for the formation of 1-pentene are in good agreement with those obtained by Tsang, our values being lower by a factor 1.2 to 2.

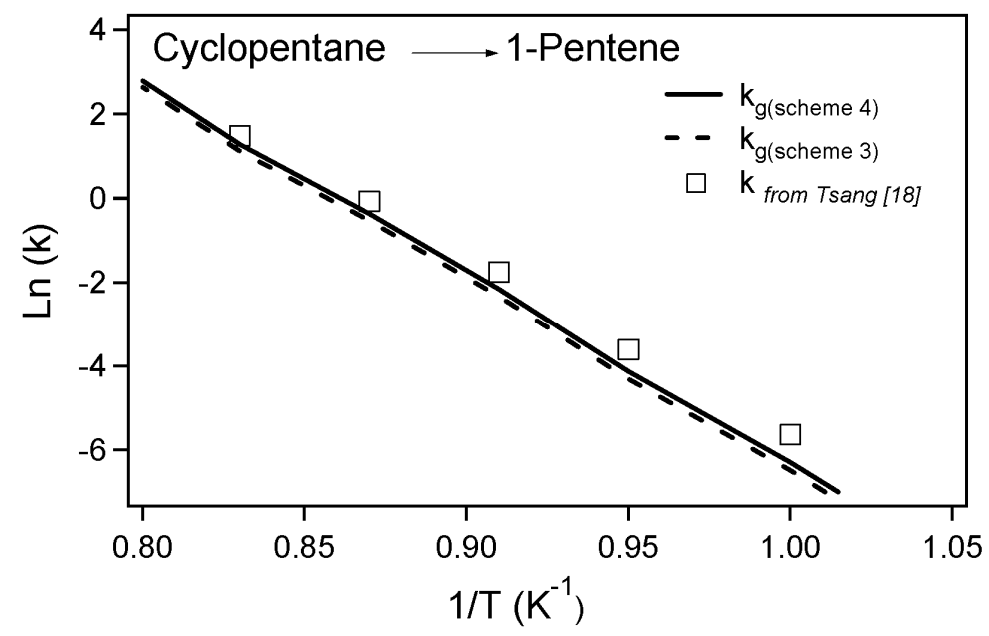

Figure 2. Comparison between calculated and experimental rate constant for the global reaction $cC_5H_{10}$ → 1-pentene

Concerning the influence of rotational hindrance in the formation of 1-pentene, scheme 3 leads to lower values than scheme 4 over the temperature range (28 % at 600K and 8% at 2000K). Calculations performed by considering the different conformations for biradical •$C_5H_{10}$• permit to obtain results closer to experiment. Moreover, the effect of rotational hindrance is slightly greater in the case of cyclopentane than cyclobutane.

**Cyclohexane**

Only a few numbers of studies have been performed on the unimolecular initiation mechanism of cylohexane [19]. Scheme 5 shows the results for the ring opening of c-$C_6H_{12}$ and subsequent reactions. As before, in this scheme, we only consider the lowest Gibbs free energy conformers of the biradicals. Table 5 summarizes the computed rate parameters.

At high temperature, the concentration of the twist boat conformation cannot be neglected and the equilibrium constant $K_{eq}$, corresponding to the reaction:

$$c\text{-}C_6H_{12}\ (\text{chair}) \rightleftharpoons c\text{-}C_6H_{12}\ (\text{boat})\ ,$$

has been fitted in the 600 - 2000 K:

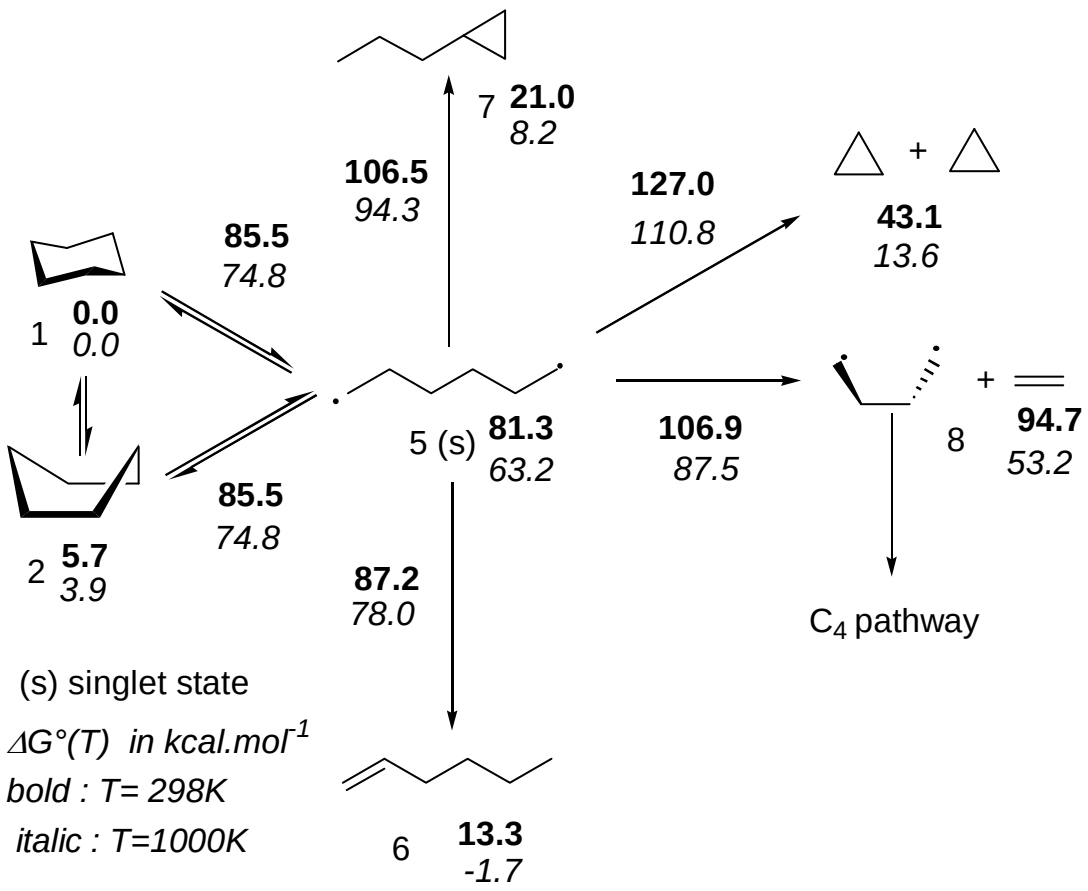

Scheme 5. Mechanism obtained for the ring opening of cyclohexane and for conformers of lowest energies.

Table 5. Rate parameters for the unimolecular initiation of cyclohexane at P=1 atm, 600 ≤ T (K) ≤ 2000 K and related to scheme 5.

| | $k_{1-5}$ | $k_{2-5}$ | $k_{5-6}$ | $k_{5-7}$ | $k_{5-8}$ |
|---|---|---|---|---|---|
| log(A $s^{-1}$) | 21.32 | 20.11 | 2.46 | -1.33 | 10.40 |
| n | -0.972 | -0.785 | 2.569 | 3.800 | 0.994 |
| $E_a$ (kcal/mol) | 92.63 | 85.77 | 1.42 | 17.22 | 25.75 |

Two different TSs have been found for the ring opening of cyclohexane depending on its initial boat or chair structure, the TS corresponding to the former conformation being 2.4 kcal.mol$^{-1}$ lower in Gibbs free energy. Since activation energy for ring opening is much higher than the energy involved in boat/chair conformation change, we conclude that only the lowest TS should be taken into account in the kinetic scheme. Actually, the distinction between chair and boat conformation is not fundamental in order to derive accurate thermodynamic properties. Nevertheless, we have taken both conformations into account with the aim of describing the detailed kinetic mechanism.

The Gibbs free energy of activation obtained for the ring opening of cyclohexane (85.5 kcal.mol$^{-1}$, at T = 298 K) is higher than that obtained for cyclobutane and close to the value found for a linear alkane. This is consistent with an unstrained structure in cyclohexane. The main way of decomposition for the biradical 5 is the formation of 1-hexene, as mentioned by Tsang [19]. Indeed, this process involves a slightly constrained transition state for H-abstraction (six-member ring) with a low activation energy of 1.8 kcal.mol$^{-1}$ (*vs* 17.6 kcal.mol$^{-1}$ for c-$C_4H_8$ and 6 kcal.mol$^{-1}$ for c-$C_5H_{10}$). Owing to this very low activation energy, other routes for •$C_6H_{12}$• decomposition are unlikely. An remark can be made for β-scission leading to the formation of •$C_4H_8$• and $C_2H_4$ (reaction 5→8). In the experimental study performed by Tsang [19], no cyclobutane was detected what could implicitly be explained by a very low reaction rate for •$C_4H_8$• formation compared to 1-hexene. Analyzing Scheme 5 shows that the ratio between the disproportionation leading to 1-hexene (reaction 5→ 6) and the β-scission (reaction 5→ 8) is about 120 at 1000 K. Moreover, the decomposition of the biradical •$C_4H_8$• in two ethylene molecules is very fast compared to the cyclization reaction, as shown in Scheme 1.

Let us now consider the effect of conformers of •$C_6H_{12}$• that was ignored in Scheme 5. The activation energy involved in the formation of 1-hexene from the biradical •$C_6H_{12}$• is quite low (1 kcal.mol$^{-1}$) and might be competitive with rotational barriers. Accordingly, we developed a detailed mechanism for the formation of 1-hexene involving the different conformations of the biradical (Scheme 6).

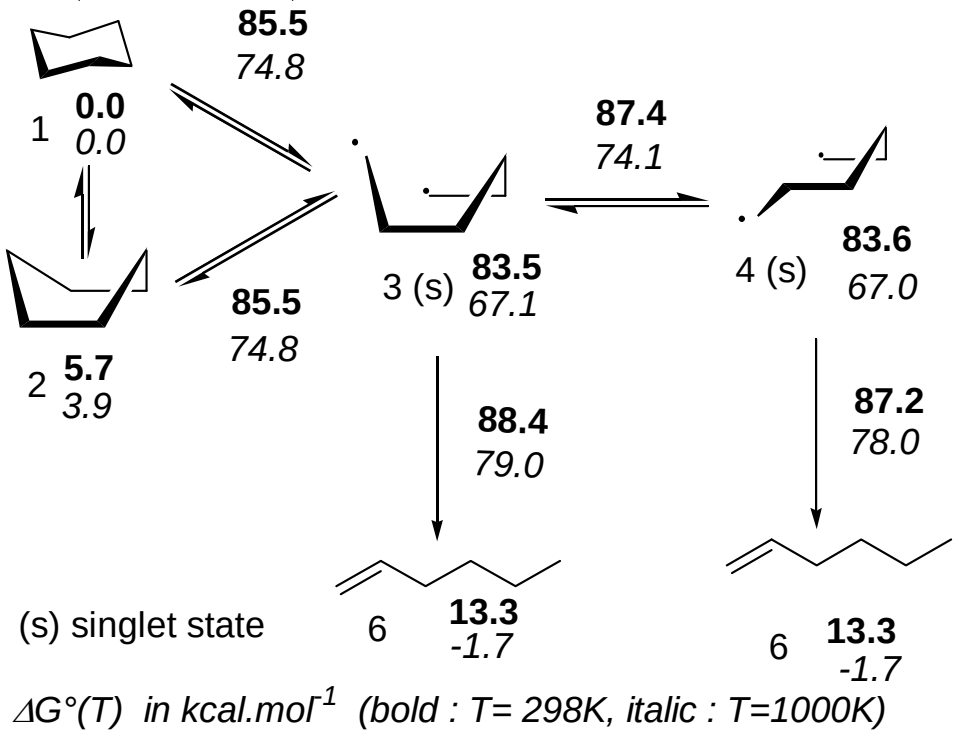

Scheme 6: Detailed mechanism of 1-hexene formation from the ring opening of cyclohexane and obtained by considering different conformers of $C_6$ biradicals.

It is worth noting that more conformers have been found for •$C_6H_{12}$• but the formation of 1-hexene from these biradicals needed an internal rotation leading to one of the two conformers considered in scheme 6. Moreover the routes of decomposition for these conformers have been all found energetically unfavourable compared to that leading to 1-hexene.

We performed QSSA on biradical 5 (Scheme 5) and biradicals 3 and 4 (Scheme 6). Rate parameters obtained for temperatures ranging from 600K to 2000K are presented in Table 6

Table 6: Rate parameters for the global reactions c$C_6H_{12}$ → 1-hexene at P=1 atm, 600 ≤ T (K) ≤ 2000 K and related to Schemes 5 and 6.

| | $k_{g(1-6)}^{scheme\ 7}$ | $k_{g(1-6)}^{scheme\ 8}$ |
|---|---|---|
| log A ($s^{-1}$) | 20.45 | 20.29 |
| n | -0.685 | -0.639 |
| $E_a$ (kcal/mol) | 93.01 | 94.52 |

Figure 3 shows the comparison of the values calculated from both schemes (Table 6) and those obtained from the rate constant proposed by Tsang [19] in the range of validity of Tsang's study.

The agreement between $k_{g(1-6)}^{scheme8}$ and the rate constant proposed by Tsang is rather satisfactory since their ratio lies between 2 and 2.6. Moreover, by neglecting rotational hindrance (Scheme 5), the global rate constant is overestimated by a factor 2 compared to scheme 6, around 1000 K. This difference can be explained by the low activation energy involved in the formation of 1-hexene compared to rotational barrier, but also by entropic effects due to the difference between entropy of the "linear" biradical (biradical 5) and biradicals 3 and 4. This results shows that by considering only the lowest energy biradical conformation, the global rate constant is over-estimated.

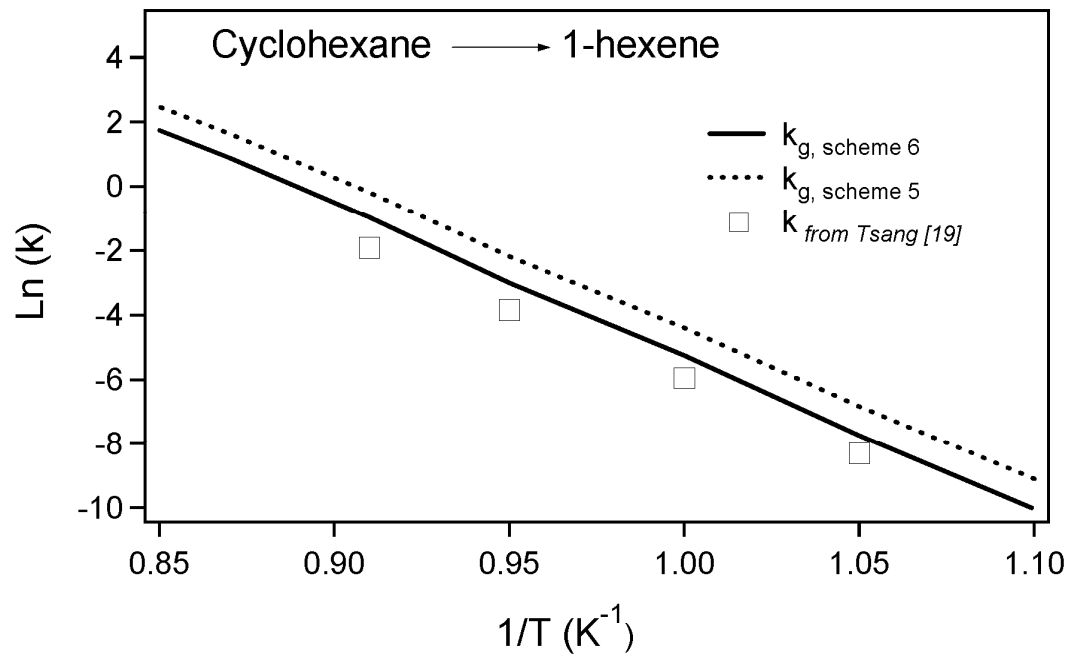

Figure 3: comparison between calculated rate constant and experimental data [19] for the global reactions c-$C_6H_{12}$ → 1-hexene.

## Conclusions

In this study, the ring opening of the most representative cyclic alkanes have been extensively explored by means of quantum chemistry. All the possible elementary reactions have been investigated from the biradicals yielded by the initiation steps.

For all reactions, the transition state theory allowed to calculate the rate constant. Thanks to the Quasi Steady State Approximation applied to the biradicals, rate constants have been calculated for the global reactions leading directly from the cyclic alkane to the molecular products. These values are in a rather good agreement with the literature. The main reaction routes are the decomposition to two ethylene molecules in the case of cyclobutane and the internal disproportionation of the biradicals yielding 1-pentene and 1-hexene in the case of cyclopentene and cyclohexane, respectively. An interesting fact highlighted in this work is the role of the internal rotation hindrance in the biradical fate. Whereas the energy barriers between conformers are usually of low energy in comparison to the reaction barriers, all the energies are close in this case and taking the rotations between the conformers into account changes the global rate constant especially for the largest biradicals.

## Acknowledgment

The Centre Informatique National de l'Enseignement Supérieur (CINES) is gratefully acknowledged for allocation of computational resources